\documentclass[a4paper,11pt]{article}

\usepackage{jcappub} 

\usepackage[T1]{fontenc} 

\title{\textbf{Photon-axion mixing within the jets of Active Galactic Nuclei and prospects for detection}} 

 \author{J. Harris}
 \author{and P. M. Chadwick}
 \affiliation{Durham University, \\Department of Physics, South Road, Durham DH1 3LE, UK}

\emailAdd{j.d.harris@durham.ac.uk}
\emailAdd{p.m.chadwick@durham.ac.uk}

\abstract{
Very high energy $\gamma$-ray observations of distant active galactic nuclei (AGN) generally result in higher fluxes and harder spectra than expected, resulting in some tension with the level of the extragalactic background light (EBL).  If hypothetical axions or axion-like particles (ALPs) were to exist, this tension could be relieved since the oscillation of photons to ALPs would mitigate the effects of EBL absorption and lead to softer inferred intrinsic AGN spectra.  In this paper we consider the effect of photon-ALP mixing on observed spectra, including the photon-ALP mixing that would occur within AGN jets.  We then simulate observations of three AGN with the Cherenkov Telescope Array (CTA), a next generation $\gamma$-ray telescope, to determine its prospects for detecting the signatures of photon-ALP mixing on the spectra.  We conclude that prospects for CTA detecting these signatures or else setting limits on the ALP parameter space are quite promising.  We find that prospects are improved if photon-ALP mixing within the jet is properly considered and that the best target for observations is PKS~2155-304.
}

\begin{document}
\maketitle
\flushbottom

\section{Introduction}
As very high energy (VHE; roughly $50\rm~GeV$-$100\rm~TeV$) $\gamma$-rays from active galactic nuclei (AGN) traverse intergalactic space they suffer absorption with low energy photons from the diffuse extragalactic background light (EBL) \citep{Gerasimova62}.  Higher energy $\gamma$-rays experience more EBL absorption and so the observed $\gamma$-ray spectrum after absorption is softened.  The intrinsic VHE $\gamma$-ray spectrum of an observed AGN therefore needs to be inferred by removing the EBL effects (see e.g. Ref~\cite{Aharonian06}).  The amount of EBL absorption is currently uncertain, with several models of varying levels in the literature.  However, even for lower-level models of the EBL, the unusually hard $\gamma$-ray spectrum inferred for some distant AGN can be difficult to account for given current understanding of $\gamma$-ray emission mechanisms \citep{Lefa11}.  This issue could be resolved if the hypothetical axion or axion-like particle (ALP) exists (e.g. Refs~\cite{Csaki03}, \cite{DeAngelis07}, \cite{Horns12}).  Oscillation of photons to ALPs in magnetic fields could mitigate the amount of EBL absorption meaning that the intrinsic spectra of AGN would be softer than previously inferred.

In this paper we consider the effect on $\gamma$-ray spectra of photon-ALP mixing in the magnetic fields both surrounding the AGN source \textit{and} within the jet of the AGN itself.  We consider the prospects of the Cherenkov Telescope Array (CTA), a next generation ground based $\gamma$-ray telescope, detecting the effects of photon-ALP mixing in AGN spectra. 

\subsection{Extragalactic Background Light and its effects on AGN spectra}
The EBL is the diffuse light present throughout the universe.  The presence of this light means that the universe has an optical depth to VHE $\gamma $-rays since they can annihilate with the lower energy EBL photons.  The EBL photons that contribute to this process are dominated by infrared photons, with optical-UV photons also playing a significant role \citep{Aharonian06}.  These photons originate primarily from stars, both via direct emission and via re-emission after absorption by dust \citep{Finke10}.  Since the number of stars and rate of star formation evolves along with the universe, the EBL intensity is a function of redshift as well as wavelength.  For extragalactic sources this attenuation is significant and is generally accounted for when modelling the spectra of VHE sources; however, the exact level of the EBL is unknown.  Determining the level directly from observations is challenging due to the weak and diffuse nature of the EBL, but a variety of models exist in the literature and a few of these will be discussed in more detail later.

\subsection{Axions and axion like particles}
As shown in Refs~\cite{Weinberg78}, \cite{Wilczek78}, axions are hypothetical particles extending the Standard Model which arise from the proposed explanation of the CP problem in QCD by Peccei \& Quinn \cite{Peccei77}.  If they do exist, they could have important effects in several areas of astrophysics, depending upon their mass and their coupling constant to photons, which allows photons to oscillate into axions and vice versa in the presence of a magnetic field.  The coupling constant is a function of axion mass, but for the more generic ALPs this need not be the case.  ALPs can play an important role in the cooling of white dwarf stars (e.g. Ref~\cite{Isern10}), in the composition of dark matter (e.g. Ref~\cite{Park12}) and in the amount of attenuation $\gamma $-rays undergo due to interaction with the EBL (e.g. Ref~\cite{Simet08}).

It has been noted several times in the literature that a proportion of the emitted $\gamma $-rays may oscillate into ALPs in the magnetic field surrounding an extragalactic source and these particles would no longer be attenuated by the EBL as they propagate through intergalactic space.  A proportion of the ALPs could then oscillate back into $\gamma $-rays in the Milky Way's magnetic field and in this way provide a boost to the measured $\gamma $-ray spectrum (see e.g. Refs~\cite{Simet08}, \cite{Horns12}). Such an effect could explain anomalies that have been noticed in the VHE spectra of distant AGN \citep{HornsMeyer12}.

It is not always taken into account that if $\gamma $-rays can oscillate into ALPs in the magnetic field surrounding the source then they must also be subject to oscillations into ALPs \textit{within} the source, since the plasma jet in which $\gamma $-ray emission originates is universally accepted to have a magnetic field.  Photon-ALP mixing within the source was considered by S\'{a}nchez-Conde et al. \cite{Sanchez-Conde09}.  However, only the magnetic field of the small ($\leq 0.3\rm~pc$) region around the $\gamma $-ray emission region was considered, and not the magnetic field of the rest of the jet through which the $\gamma $-rays must propagate.  Photon-ALP mixing within an AGN jet was also discussed briefly by Hochmuth \& Sigl \cite{Hochmuth07}, who considered conversion within a slightly larger (parsec scale) region with a homogeneous magnetic field.  Tavecchio et al. \cite{Tavecchio12} considered mixing in a jet of length $6.7\rm~kpc$ as a model for the FSRQ PKS~1222+216 and Mena \& Razzaque \cite{Mena13} consider mixing in the jet of this object and one other FSRQ object assuming a jet length of $3.2\rm~kpc$ in each case.

In this paper, we consider photon-ALP mixing through jets ranging in length between $0.1\rm~kpc$ and $100\rm~kpc$, with a magnetic field throughout that decreases steadily from a maximum value at the $\gamma $-ray emission region and has random orientations for different cells.  We simulate the propagation of an initial beam of $\gamma $-ray photons and calculate the emitted spectrum for various jet sizes and consider how probable it is for the effects of photon-ALP mixing to be observed, both in the case where mixing occurs in the jet and where mixing occurs only outside the jet.  This paper is structured as follows: in Section~2 we discuss the size of AGN jets, the site of the $\gamma$-ray emission region, and the jet model used in this paper.  In Section~3 we present the equations that govern photon-ALP mixing and apply these equations to our jet model to calculate the emitted $\gamma$-ray spectrum.  In Section~4 we discuss the prospects for CTA detecting the effects of photon-ALP mixing in $\gamma$-ray spectra.  We discuss the models used for the EBL and for the Milky Way's magnetic field, present our target objects, describe our simulations of them with CTA and present the results of these simulations.  A short summary of the paper is given in Section~5.

\section{AGN Jets}
By definition, blazars are viewed at a small angle to the jet.  While the beaming effects of looking down the jet can create very bright sources it makes determining the jet size and emission region of individual sources difficult.  Some inferences can be made from VLBI observations of nearby AGN viewed at large angles to the jet.  For Fanaroff-Riley~I type galaxies, collimated jets of a few $\rm{kpc}$ are common \citep{Stawarz03} and jets of several hundred $\rm{kpc}$ are known \citep{Blundell01}.  The $\gamma $-ray emission region appears to be located relatively close to the base of the jet, at distances $<200\rm~pc$ (see Ref~\cite{Cheung07} for VLBI observations and Ref~\cite{Tavecchio13} for arguments from$\rm~GeV$ energy observations of FSRQ type blazars).  This suggests that, in general, the distance from the emission region to the termination of the jet $\Delta Z$ is $>1\rm~kpc$.  Disruption of the jet close to its base means that $\Delta Z$ could be significantly smaller in some objects, although there is less chance of this being the case in brighter objects \citep{Blundell01}.  These considerations lead us to consider several values of $\Delta Z$ in our work: $0.1$, $1$, $10$ and $100\rm~kpc$.

\subsection{Modelling}\label{sec:AGNmod}
We consider propagation of $\gamma $-rays through a jet with oscillation to ALPs along its length.  The $\gamma $-ray spectrum is based on the measurement of the BL Lac object 1ES~1101-232 and the underlying physical parameters are based on the subsequent modelling \citep{Aharonian07}.  At the emission region we therefore have electron density $K_{0}=900\rm~cm^{-3}$ and magnetic field strength $B_{0}=1\rm~G$, which is consistent with the average value for the magnetic field found for the emission region of $\gamma $-ray loud BL Lacs in the study by Ghisellini et al. \cite{Ghisellini10}.  The $\gamma $-ray flux $F$ at energy $E_{\gamma}$ is therefore given by
\begin{equation}
F(E_{\gamma})=5.6 \times 10^{-13} \left(\frac{E_{\gamma}}{1\rm~TeV}\right)^{-2.94}\rm~ph~cm^{-2}~s^{-1}~TeV^{-1},
\end{equation}
where $1\rm~MeV \leq E_{\gamma} \leq 10\rm~TeV$.  This is a considerable extrapolation from the measurement in Ref~\cite{Aharonian07}, but this is not important for our current purposes where we wish only to test the effects of photon-ALP oscillation over a wide energy range and are not too concerned about the actual spectral shape at this stage.  The extrapolation below a few tens of $\rm{GeV}$ would be below the sensitivity threshold for CTA, but we include this in our modelling of the jet as effects in this range could be detected by other instruments e.g. the \textit{Fermi}-Large Area Telescope (LAT).

The jet model used in this paper is based loosely on the fundamental model presented by Blandford \& K\"onigl \cite{Blandford79}.  The jet is assumed to have a constant opening angle $\phi=0.1^{\circ}$ relating the jet radius $R$ to distance from the base of the jet $Z$ as follows:
\begin{equation}
R(Z)=\phi \cdot Z.
\end{equation}
The $\gamma $-ray emission region is set at a distance $Z_{0}=1\rm~pc$ from the base of the jet and the jet terminates at a distance $Z_{max}=Z_{0}+\Delta Z$ where we take $\Delta Z$ as $0.1$, $1$, $10$ and $100\rm~kpc$.  The electron density and the magnetic field strength have their maximum values at the emission region and then decrease up the jet, as we shall see.

Conserving particle number up the widening jet, we have
\begin{equation}
K(Z) \propto \frac{1}{Z^{2}}.
\end{equation}

Following Heinz \& Sunyaev \cite{Heinz03} we assumed that the kinetic jet power is proportional the the mass accretion rate, $\dot{m}$ (see also \cite{Falcke95}).  Following Gardner \& Done \cite{Gardner13}, we further assume that every unit time a constant fraction of kinetic energy per unit volume, $\frac{KE}{V}$, is converted into magnetic energy.  These assumptions give us a magnetic energy density of
\begin{equation}
U_{B} \propto \frac{KE}{V} \propto \frac{\dot{m}}{\pi r^{2} \beta c},
\end{equation}

where $\pi r^{2} \beta c$ is the volume swept out per unit time by material passing up the jet.  Since $r \propto Z$ and $U_{B} = \frac{B^2}{8\pi}$ we see that
\begin{equation}
B(Z) \propto \frac{1}{Z}.
\end{equation}

Following S\'{a}nchez-Conde et al. \cite{Sanchez-Conde09} we take the coherence length of the magnetic field (the distance over which the direction of the magnetic field is constant) to be $3 \times 10^{-3}\rm~pc$.  We begin with a pure photon beam and for $20$ energy bins, evenly spaced on a logarithmic scale, calculate the fraction of photons oscillating into ALPs and vice versa when travelling over cells of one coherence length.

This jet model, although simplistic, covers all the salient points for our purposes.  That is, the $\gamma $-rays are emitted relatively close to the base of the jet and must propagate large distances and many coherence lengths of a magnetic field of changing strength.  As discussed later, our conclusions should not be affected if the $\gamma $-ray emission region has a larger magnetic field or particle density than its surroundings, such as for blob-in-jet \citep{Katarzynski01} or magnetic reconnection \citep{Lyubarsky10} scenarios.

\section{Photon ALP mixing}
\subsection{Mixing Equations}\label{sec:MixingEq}
Following Refs~\cite{Raffelt88}, \cite{Hochmuth07} and \cite{Horns12}, we take the Lagrangian for coupling between the electromagnetic field and the axion field as 
\begin{equation}
\mathcal{L}=g\underline{E} \cdot \underline{B}a,
\end{equation}
where $g$ is a coupling constant, $\underline{E}$ and $\underline{B}$ are the electric and magnetic fields respectively, and $a$ is the axion field of mass $m_{a}$.  Following \cite{Horns12} we take $g=5 \times 10^{-11}\rm~GeV^{-1}$ and $m_{a}=1 \times 10^{-8}\rm~eV$, which are within the constraints determined by the CAST experiment \citep{Andriamonje07}.

With this Lagrangian, over a length $s$ of plasma with approximately constant frequency and magnetic field strength, the probability for a photon of energy $E_{\gamma}$ to convert into an ALP (or vice versa) is\footnote{There is some disagreement as to the correct form of equation~\ref{P0}: see Ref~\cite{arXivnote} and references therein.  If the choice in this paper is incorrect then $P_{0}$ would increase by a factor of $2$.  This would not have a significant impact on our conclusions given the discussion in the next section.}
\begin{equation}\label{P0}
P_{0}=2\left(\frac{\Delta_{B}}{\Delta_{osc}}\right)^{2}\sin ^{2}\left(\frac{ \Delta_{osc}\cdot s}{2}\right).
\end{equation}

Here $\Delta_{B}$ is a term relating to the transverse strength of the magnetic field $B_{T}$,
\begin{equation}\label{deltaB}
\Delta_{B}=7.6 \times 10^{-2} \left(\frac{g}{5 \times 10^{-11}\rm~GeV^{-1}}\right)\left(\frac{B_{T}}{10^{-6}\rm~G}\right)\rm~kpc^{-1},
\end{equation}
and $\Delta_{osc}$ is the oscillation wave number,
\begin{equation}\label{deltaOsc}
\Delta_{osc}=\Big( \left(\Delta_{CM} + \Delta_{pl} - \Delta_{a}\right)^{2} + 4\Delta_{B}^{2}\Big)^{\frac{1}{2}}\rm~kpc^{-1},
\end{equation}
$\Delta_{CM}$ is the Cotton-Mouton term,
\begin{equation}\label{deltaCM}
\Delta_{CM}=-4 \times 10^{-6}\left(\frac{B_{T}}{10^{-6}\rm~G}\right)^{2}\left(\frac{E_{\gamma}}{\rm~TeV}\right)\rm~kpc^{-1},
\end{equation}
$\Delta_{pl}$ is the plasma term,
\begin{equation}
\Delta_{pl}=1.1 \times 10^{-10}\left(\frac{E_{\gamma}}{\rm~TeV}\right)^{-1}\left(\frac{K}{10^{-3}\rm~cm^{-3}}\right)\rm~kpc^{-1},
\end{equation}
where $K$ is the electron density of the plasma, and $\Delta_{a}$ is a term relating to the axion mass,
\begin{equation}\label{deltaA}
\Delta_{a}=7.8 \times 10^{-3}\left(\frac{m_{a}}{10^{-8}\rm~eV}\right)^{2}\left(\frac{E_{\gamma}}{\rm~TeV}\right)^{-1}\rm~kpc^{1}.
\end{equation}

For the values relevant to this work, Faraday rotation effects can be neglected \citep{Ganguly09}.
When $P_{0} \simeq 1$, photon-ALP conversion is saturated, since for every photon converting into an ALP there is also an ALP converting to a photon.  Over energies where this strong mixing occurs, the intrinsic spectral shape is unaltered but the photon flux will drop by $\frac{1}{3}$ due to their conversion to axions (for an unpolarised photon beam as is assumed here).

We can see from equation~\ref{P0} that there are two conditions to be met for strong photon-ALP mixing to take place.  The first condition is that the argument of the $\sin^{2}$ term is non-negligible,
\begin{equation}\label{req2}
\frac{ \Delta_{osc}\cdot s}{2}\gg0.
\end{equation}
$\Delta_{osc}$ can only be consistently small if the strength of the magnetic field is small\footnote{It is of course small if $B_{T}$ is small solely due to the angle of photon propagation with respect to the magnetic field, but on average this will not be the case.}.  In this case,
\begin{equation}
\Delta_{osc} \simeq \sqrt{ \left(\Delta_{pl} - \Delta_{a}\right)^{2}}.
\end{equation}
$\Delta_{pl}$ is a function of the electron density which, unless $\Delta Z$ is very small, varies over several orders of magnitude. $\Delta_{a}$ on the other hand is a function only of the ALP mass (which is fixed) and photon energy.  For this work, $1\rm~MeV \leq E_{\gamma} \leq 10\rm~TeV$ which leads to $3\times 10^{11} \leq \Delta_{a} \leq 3 \times 10^{17}\rm~kpc^{-1}$.  Given the coherence path length adopted in this work of $s=3 \times 10^{-3}\rm~pc$, equation~\ref{req2} is easily satisfied for much of the jet.

The second condition for strong mixing is
\begin{equation}\label{req1}
2\left(\frac{\Delta_{B}}{\Delta_{osc}}\right)^{2} \simeq 1,
\end{equation}

$\Delta_{B}$ is dependent on the strength of the magnetic field and $\Delta_{osc}$ is  dependent on both this and the electron density.  Unless $\Delta Z$ is very small, these quantities vary over orders of magnitude through the jet, and for much of the jet equation~\ref{req1} will be satisfied.

At the very highest values of photon energy, $E_{\gamma}$, the Cotton-Mouton term described in equation~\ref{deltaCM} becomes important.  Unless $B_{T}$ is small, $\Delta_{CM} \gg \Delta_{B}$ and will dominate $\Delta_{osc}$, meaning that equation~\ref{req1} will no longer be satisfied.  If $\Delta Z$ is small, conservation of magnetic energy dictates that $B_{T}$ is generally large and so photon-ALP mixing is suppressed at high energies.

As pointed out in Ref~\cite{DeAngelis08}, it can be useful to define a critical energy
\begin{equation}
 E_{c}= \frac{5\times10^{-2} |m_{a}^{2}-\omega_{pl}^{2}|}{\left(10^{-8}\rm~eV\right)^{2}} \
\left(\frac{10^{-6}\rm~G}{B_{T}}\right) \left(\frac{5 \times 10^{-11}\rm~{GeV}^{-1}}{g}\right)\rm~TeV,
\end{equation}

where $\omega_{pl}=-2E\Delta_{pl}$.  This allows us to rewrite equation~\ref{P0} as
\begin{equation}
P_{0}=\frac{1}{1+\left(\frac{E_{c}}{E}\right)^{2}}\sin^{2}\left(\frac{\Delta_{osc} \cdot s}{2}\right).
\end{equation}
This expression neglects the Cotton-Mouton term and therefore doesn't hold at high values of photon energy.  From this expression we can see that below $E_{c}$ there is little photon-ALP oscillation and above $E_{c}$ there is potentially significant photon-ALP oscillation, a fact we shall refer to later in the paper.

\subsection{Mixing within AGN jets}
\begin{figure*}
\includegraphics[scale=0.95]{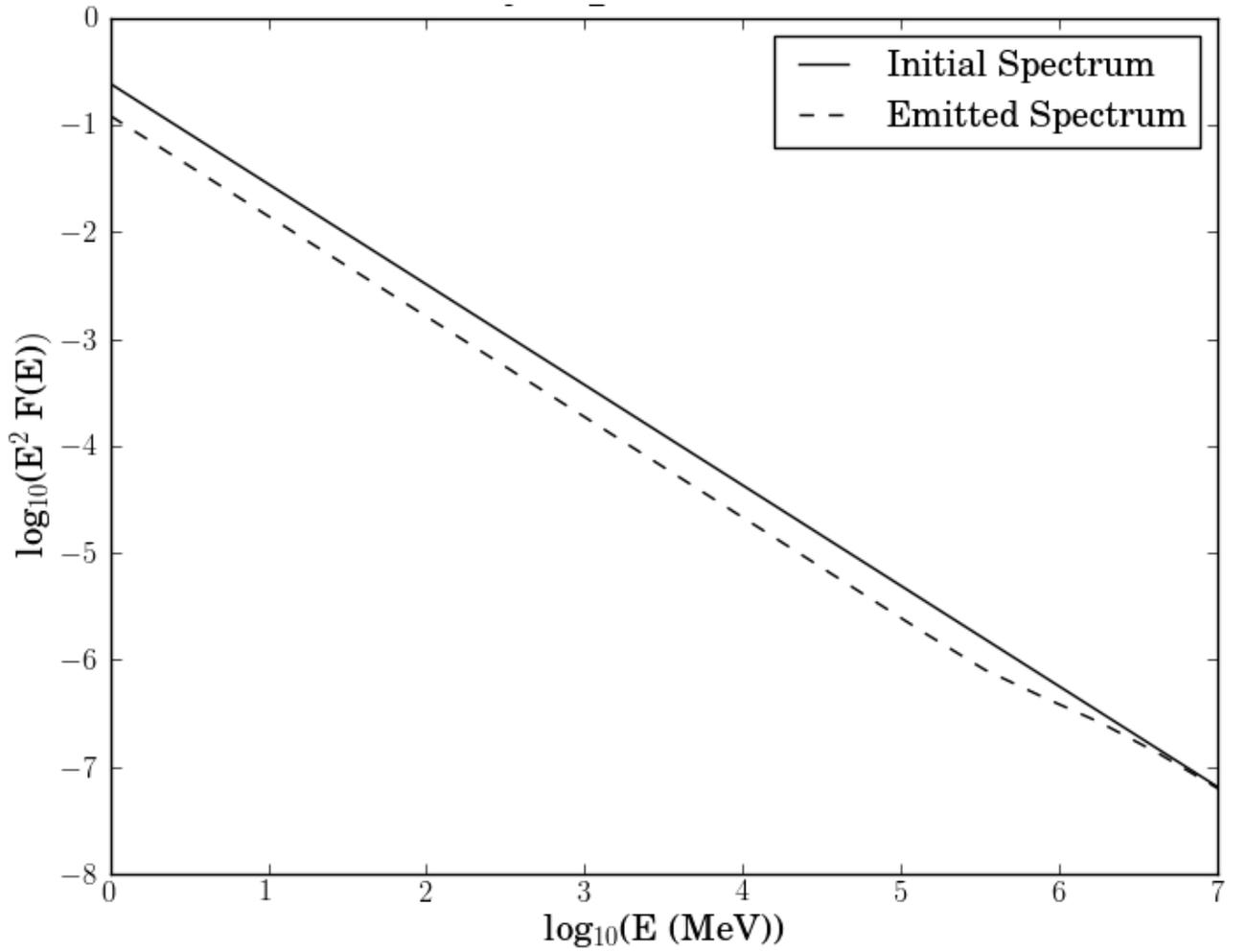}

\caption{\label{fig:z01}The intrinsic photon spectrum before any mixing with ALPs and the emitted photon spectrum after travelling 0.1~kpc from the emission region while undergoing mixing with ALPs. F is the flux as a function of E in arbitrary units.}

\end{figure*}

\begin{figure*}
\includegraphics[scale=0.95]{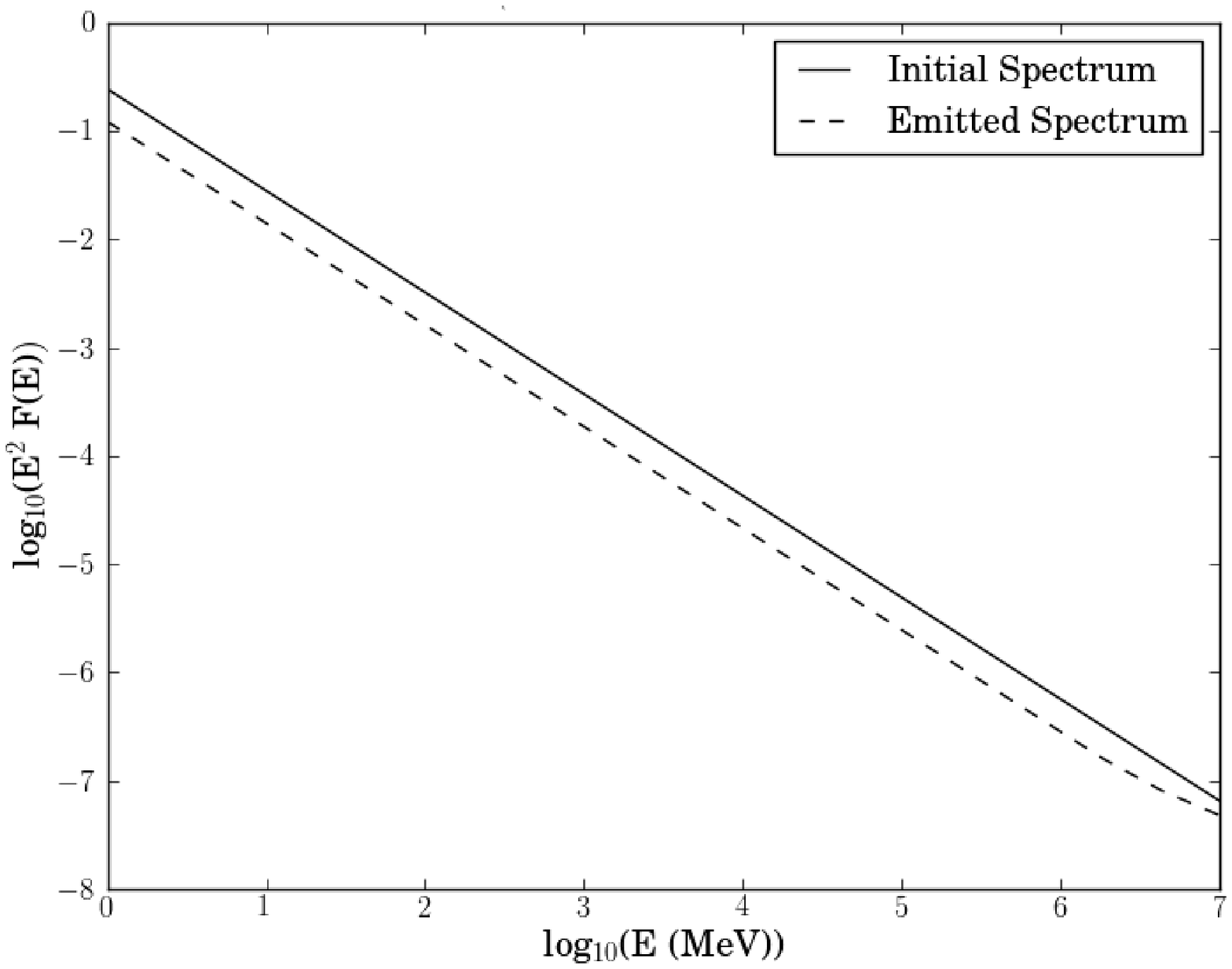}

\caption{\label{fig:z1}The intrinsic photon spectrum before any mixing with ALPs and the emitted photon spectrum after travelling 1~kpc from the emission region while undergoing mixing with ALPs. F is the flux as a function of E in arbitrary units.}

\end{figure*}

\begin{figure}
\includegraphics[scale=0.75]{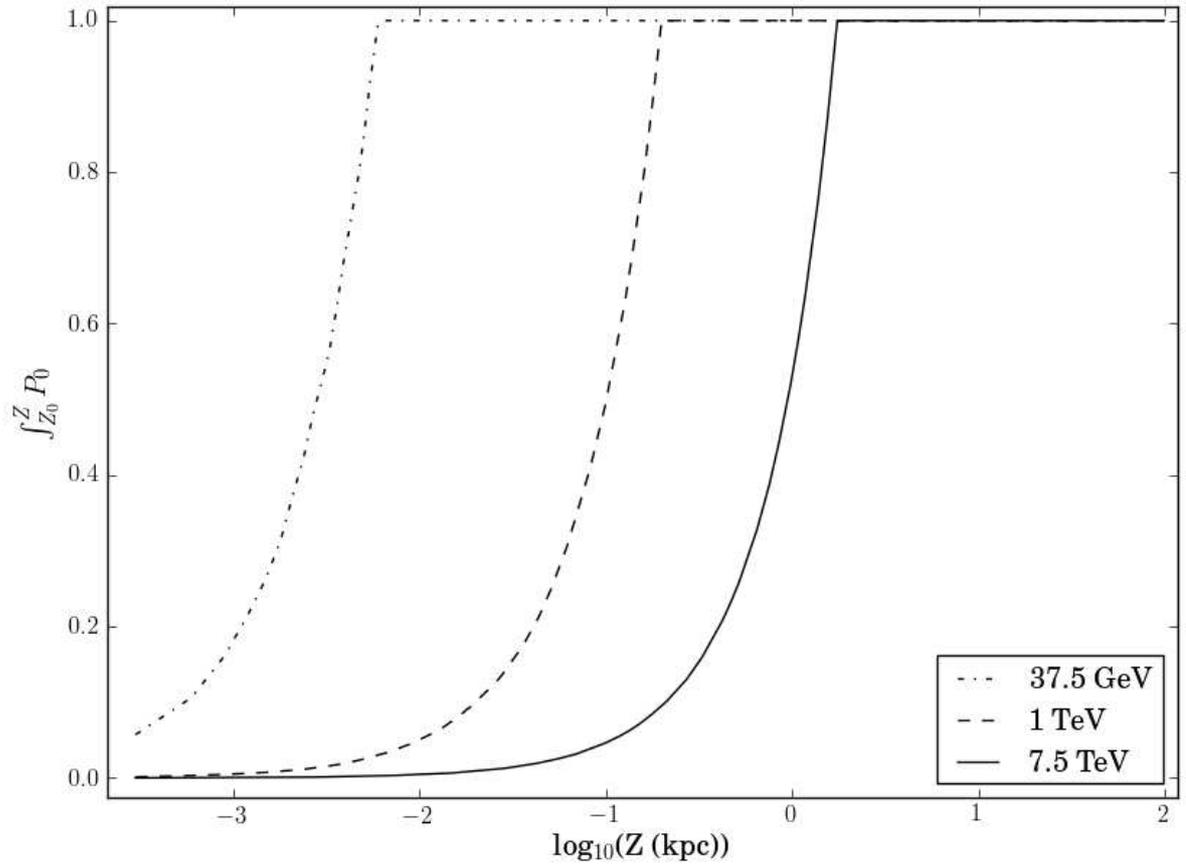}
																			   
\caption{\label{fig:Conv_Plot}The probability for (from left to right) a photon of 37.5 GeV, 1 TeV and 7.5 TeV converting into an axion at least once as a function of distance from the emission region.}

\end{figure}

At this stage we are ready to present the output of the models described in Section~\ref{sec:AGNmod}.  We found that due to the large number of cells ($>1\times 10^{4}$) the random orientations of the magnetic field in each cell had negligible effects on the results.  Figures~\ref{fig:z01} and~\ref{fig:z1} show the intrinsic and emitted spectra for AGN with different values of $\Delta Z$ and Figure~\ref{fig:Conv_Plot} shows the probabilities of photons of various energies converting into an ALP at least once, as function of distance from the emission region.  As expected from the discussion in Section~\ref{sec:MixingEq}, strong photon-ALP mixing occurs for large ranges of $E_{\gamma}$, altering the photon flux but leaving the spectral shape the same, until Cotton-Mouton effects suppress mixing at high energies.  We find that for $\Delta Z=0.1\rm~kpc$ Cotton-Mouton suppression occurs at $\approx 1.7 \rm~TeV$ resulting in a boost in the photon flux above this energy.  For larger values of $\Delta Z$, Cotton-Mouton suppression occurs at such high energies that the effect is negligible over the energy range considered in this paper.  Since no change in the spectral shape is seen below $E(\gamma) \approx 1.7 \rm~TeV$ instruments operating below this energy, e.g. the \textit{Fermi}-LAT, would not be expected to detect any effects due to photon-ALP mixing within the jet for the ALP parameters assumed here.

Before moving on we should note that when considering propagation over many coherence lengths, as we are here, the conversions should properly be treated as a three-dimensional problem. Relative to the direction of propagation, the photons can have two directions of polarisation.  As we saw, photon-ALP oscillation is only dependent on the \textit{transverse} strength of the magnetic field so each polarisation state interacts differently.  The axion field comprises the third degree of freedom.  In this paper we neglect the polarisation of photons and the equations in Section~\ref{sec:MixingEq} are derived from a two-dimensional treatment.  This will limit the accuracy of the results, but since the photon-ALP mixing probability is close to unity when outside the Cotton-Mouton regime and close to zero when inside, with a very sharp transition, it will not affect our conclusions.  Even if this were not the case, it would be expected that the results would be correct to within an order of magnitude, which is adequate given the larger uncertainties being dealt with such as the ALP mass and the length of the AGN jet.

\section{Prospects for observing photon-ALP mixing effects}
\subsection{The Cherenkov Telescope Array}
The Cherenkov Telescope Array (CTA) is a next-generation $\gamma $-ray telescope that will operate in the VHE regime from a few tens of$\rm~GeV$ to over $100\rm~TeV$ \citep{Acharya13}, extending the energy range of current instruments by roughly a factor of ten in each direction.  Additionally, CTA promises to be able to measure the spectra of objects with unprecedented accuracy over this energy range.  There will be two arrays, one in the Northern hemisphere and a larger one in the Southern hemisphere.  The exact layout is as yet undecided and several proposed layouts have been characterised through extensive Monte Carlo design studies including calculating the effective area for numerous energy bins \citep{Bernlohr13}. In this work we simulate observations using the results for the proposed layout~E \citep{DiPierro11} for the Southern array, which comprises of $4$ large ($\sim 24\rm~m$ primary mirror diameter) size telescopes, $23$ medium ($\sim 12\rm~m$) size telescopes, and 32 small ($\sim 4$-$7\rm~m$) size telescopes.

\subsection{Target Objects}
For this study, we choose three blazars for simulated observations which are listed in Table~\ref{ObjTable} along with their spectral parameters.  The spectrum of each object has the form
\begin{equation}
F(E_{\gamma})=k \left(\frac{E_{\gamma}}{E_{0}}\right)^{-\Gamma},
\end{equation}
where $k$ is the flux constant, $E_{0}$ is the flux normalisation and $\Gamma$ is the spectral index.

All three objects reside within a galaxy cluster and so, neglecting any conversion within the jet, it is expected that a  fraction of roughly $0.3$ of the photons convert into ALPs before traversing the IGM \citep{Horns12}.  If there is conversion within the jet then a total fraction of $\sim 0.4$ photons convert to ALPs before crossing the IGM ($\frac{1}{3}$ of photons convert to ALPs within the jet, $0.3$ of these convert back into photons in the magnetic field of the cluster, and $0.3$ of the remaining $\frac{2}{3}$ of photons convert to ALPs).  We consider both scenarios in this paper.

The first object is 1ES~1101-232.  At redshift $z=0.186$ it is relatively distant by the standards of VHE sources but is also observed to be relatively bright \citep{Aharonian07}.  Observations also show that it has a hard spectral index which leads to reasonable photon numbers at the highest energies where EBL absorption is highest and therefore any mitigation of the absorption is most easily spotted.

The second object considered is PKS~2005-489.  This object was chosen primarily because with a redshift of $z=0.071$ it is one of the closest blazars seen at VHE in the Southern hemisphere and this makes the effects of the EBL on the spectrum less model-dependent as we shall see in the following subsection.  Also, from Figure~4 of Ref~\cite{Acero10}, it can be seen that there is a slight rise in flux at a few$\rm~TeV$.  This might be construed as evidence of a flux boost from the onset of Cotton-Mouton suppression, although with current observations the effect is not statistically significant, and a simple power law fit to the data provides an adequate $\chi^{2}$.

The third object considered is PKS~2155-304, basing the spectrum on parameters in the quiescent state \cite{Abramowski10}. This object was chosen as it is one of the brightest blazars at$\rm~TeV$ energies.

The fraction of ALPs that reconvert in the Milky Way depends upon the line of sight from Earth to the object.  Our estimates, taken from Ref~\cite{Horns12}, for the fraction of photon-ALP conversion in the Milky Way's magnetic field using the model of Jansson \& Farror \cite{Jansson12} for each object are shown in Table~\ref{ObjTable}.  We also use a very low fraction of $0.05$, consistent with the lower estimate magnetic field model of Pshirkov et al. \cite{Pshirkov11}.  As we saw earlier, photon-ALP oscillation is only non-negligible above a certain critical energy.  Under the model of Jansson \& Farror \cite{Jansson12}, this energy is quite low.  In the case of PKS~2005-489 and PKS~2155-304 it is $0.016\rm~TeV$ and $0.019\rm~TeV$ respectively, below the energy threshold of CTA, and for 1ES~1101-232 it is $0.045\rm~TeV$ and only the lowest energy bin of the effective area would be below the critical energy.  Since in practice this region would be subject to quite poor energy resolution we treat this bin as also being above the critical energy.  For the model of Pshirkov et al. \cite{Pshirkov11}, with the low value of magnetic field, we calculate the critical energy as $0.2\rm~TeV$.

\subsection{EBL models}
EBL absorption can have significant effects on a spectrum, especially at high redshifts, so the choice of EBL model used could prove important. In this paper we make use of two models of the EBL.  The first is that of Kneiske \& Dole \cite{Kneiske10} which aims to provide a lower limit for the level of the EBL.  In this model certain tracers for star formation such as Lyman-$\alpha$ emission were used to infer the level of star formation as a function of redshift.  From this the EBL level was computed by fitting star formation models to the inferred lower limits.

The second model used is that of Franceschini, Rodighiero \& Vaccari \cite{Franceschini08} which provides a more typical estimate than the lower limit model.  In this model the EBL level was estimated by taking galaxy counts and attaching to each object an appropriate multiwavelength spectrum based on the galaxy type (e.g. AGN, starburst).  These spectra were modelled on local observations and evolved backwards to the appropriate redshift following a simple phenomenological model.

\begin{figure}
\includegraphics[scale=0.75]{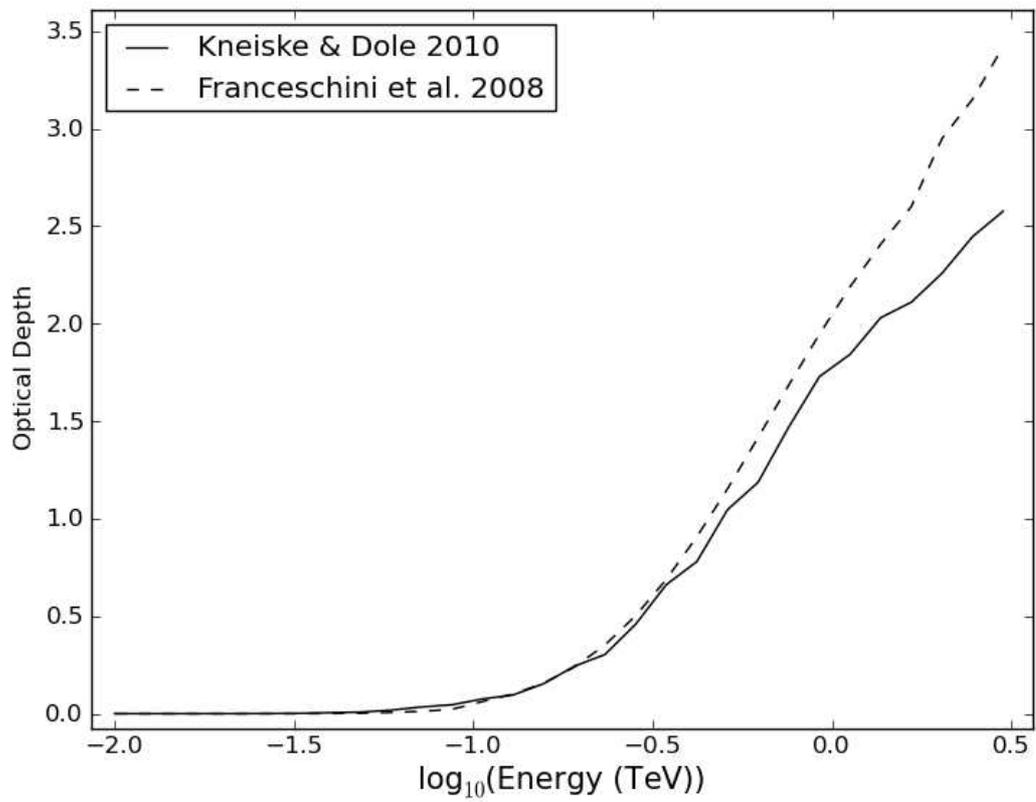}
																			   
\caption{\label{fig:OpticalDepths}Comparison of optical depths for the two EBL models used in this paper as a function of energy. These results are for the redshift of 1ES~1101-232, z=0.186.}

\end{figure}

Both models agree with the consensus that there are two peaks in the spectrum of EBL intensity, one at $\sim 1\rm~\mu m$ and another at $\sim 100\rm~ \mu m$.  However, there are significant differences between the models which become particularly apparent at high energies.  The optical depths of these two EBL models as a function of energy, at the redshift of 1ES~1101-232, are shown in Figure~\ref{fig:OpticalDepths}.

\begin{table*}
 \centering
 \footnotesize\setlength{\tabcolsep}{2.5pt}
 \begin{minipage}{140mm}
  \caption{Target objects for simulations.\label{ObjTable}}
  \begin{tabular}{@{}lcccccl@{}}
  \hline

Object & Redshift & Flux Constant &  Flux Normalisation & Spectral & Reconversion & Ref \\
Name & & $\rm{ph~cm^{-2}~s^{-1}~TeV^{-1}}$ & $\rm{TeV}$ & Index ($\Gamma$) & Frac\footnote{Reconversion of ALPs to photons in the Milky Way B-field assuming the model of Jansson \& Farror \cite{Jansson12}.} &  \\
\hline
1ES 1101-232 & 0.186 & $5.6 \times 10^{-13}$ & $1$ & $2.94$ & $0.25$ & \cite{Aharonian07} \\
PKS 2155-304  & 0.117 & $1.9 \times 10^{-12}$ & $1$ & $3.53$ & $0.60$ & \cite{Abramowski10}\footnote{Quiescent State}   \\
PKS 2005-489  & 0.071 & $1.4 \times 10^{-11}$ & $0.4$ & $3.20$ & $0.70$ & \cite{Acero10} \\

\hline

 \end{tabular}
 \end{minipage}
 \end{table*}

\subsection{Simulated Observations}
We now proceed to see how probable it is that CTA will be able to detect changes in the spectra of the target objects due to photon-ALP oscillation.  From our discussion so far, we know that the way in which the spectral shape of an object is altered by photon-ALP oscillation depends on several factors, namely: the value of $\Delta Z$, the shape of the EBL, the structure of the Milky Way's magnetic field, and whether photon-ALP oscillation occurs in the jet.

For each target we simulated $1000$ observations with CTA using the effective area curve calculated via Monte Carlo simulations of the proposed layout Array~E, described in Ref~\cite{Bernlohr13}.  In order to test the relevant factors we considered the case where Cotton-Mouton suppression is important ($\Delta Z=0.1\rm~kpc$, leading to a flux boost at $1.7\rm~TeV$), the case where Cotton-Mouton suppression is not important, and the case where there is no photon-ALP conversion within the jet at all.  Both EBL and Milky Way magnetic field models discussed in the previous sub-sections were considered.

In each case the intrinsic spectral parameters of the source were altered such that the measured spectrum would be consistent with current observations. For each simulation $50$~hours of observation and a signal-free background region (or regions) $10$~times larger than the signal region were assumed.  A simulated observation was performed as follows:
\begin{itemize}
\item The photon flux from the emission region was calculated for each energy bin.
\item To represent photon-ALP mixing within the jet and in the magnetic field surrounding the object, a fraction of this flux was converted into an ALP flux.  The fraction was based on the particular scenario: $0.4$ if there is conversion within the jet, $0.3$ if not, and no conversion in energy bins suppressed by Cotton-Mouton effects.
\item EBL absorption was applied to the photon flux based on the particular scenario, representing the beam crossing intergalactic space.
\item A fraction of photons was converted into ALPs in the Milky Way magnetic field based on the particular scenario, as shown in Table~\ref{MainTable}.  These particles were then discarded from the simulation.  A fraction of the ALP flux was then converted to photons and added to the photon flux.
\item To create a measured spectrum, the number of source events and background events for each energy bin were then drawn from a Poisson distribution about their expected values and the measured flux recorded for this bin was the number of excess events found over the expected background rate.  The expected background rate was found by drawing events for the background region from a Poisson distribution.
\end{itemize}

Two models were then fitted to the simulated spectrum.  In the first model, ALPs were not included and the flux normalisation and spectral index were the only free parameters. In the second model, the fraction of photons converting to ALPs before traversing the IGM was also a free parameter.  In each case, both the generating model and the fitted model used the same EBL model, Milky Way magnetic field model and, where appropriate, Cotton-Mouton suppression.

\begin{figure}
\includegraphics[scale=0.66]{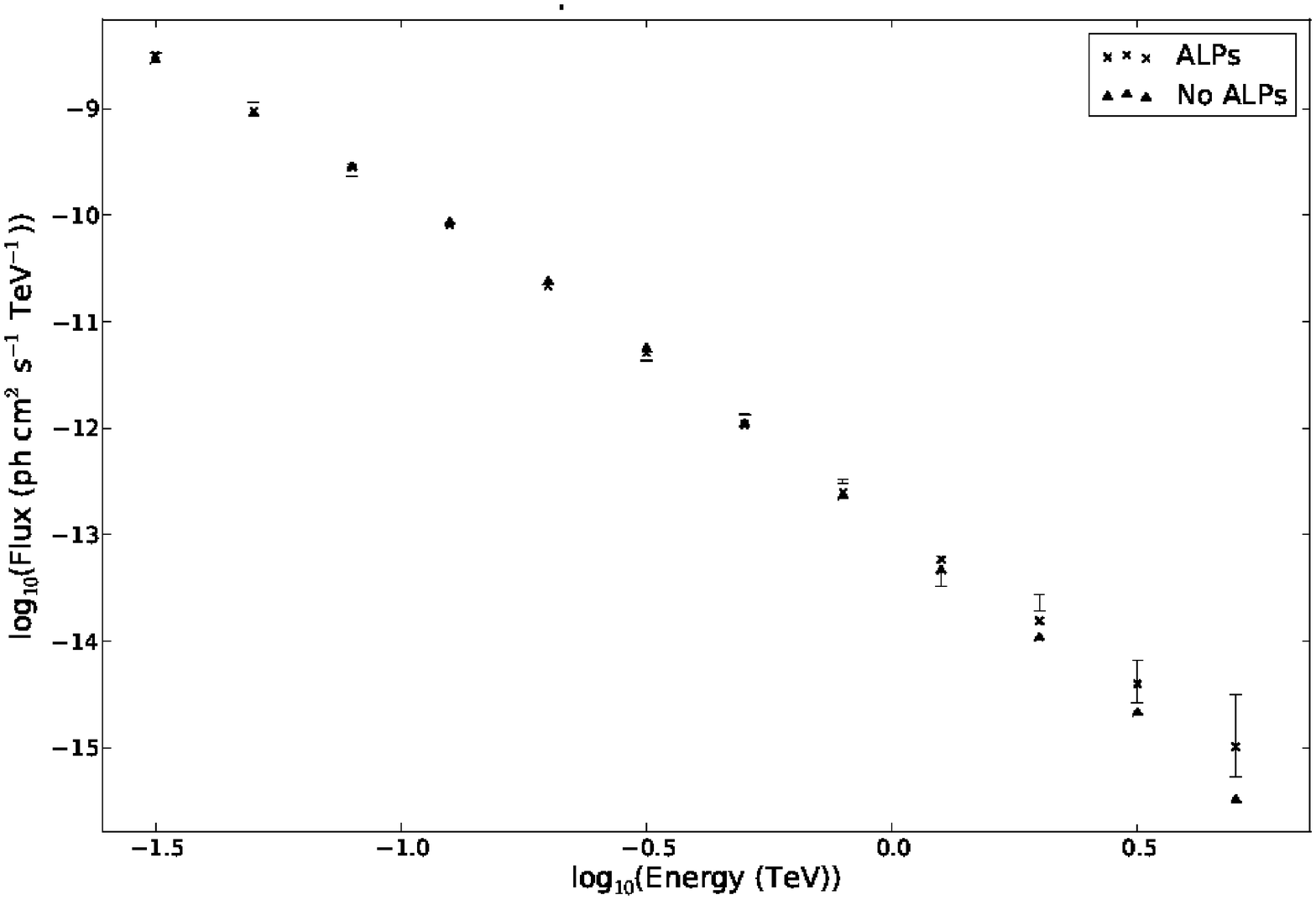}
																			   
\caption{\label{fig:SimSpec} Example of a simulated spectrum of 1ES~1101-232 where the effect of photon-ALP mixing is significant.  Error bars indicate the $1 \sigma$ statistical uncertainty on the measured photon flux in each energy bin, taken as the square root of the estimated number of events above the background.  EBL model used is that of Kneiske \& Dole \cite{Kneiske10}, with mixing occurring in the jet and in the magnetic field of the host galaxy cluster.  A model including ALPs fitted to the data is shown with crosses and a model with no ALPs fitted to the data is shown with triangles.}

\end{figure}

To see if including ALPs in the model significantly improved the fit, the likelihood of each model given the simulated data was calculated and compared using an Akaike Information Criterion ($AIC$) test.  An example of a simulated spectrum which is fitted significantly better by a model with ALPs (compared to a model without ALPs) is shown in Figure~\ref{fig:SimSpec}.

The $AIC$ value of a model $A$ is defined as

\begin{equation}
AIC_{A}=-2\ln L_{A} + 2k_{fA},
\end{equation}
where $L_{A}$ is the likelihood of model $A$ given the data and $k_{fA}$ is the number of free parameters of the model.  The difference in $AIC$ between two models $A$ and $B$ is therefore given as

\begin{equation}\label{deltaAIC}
\Delta AIC_{AB}=AIC_{A}-AIC_{B}.
\end{equation}

$\Delta AIC_{AB}$ estimates the difference in the Kullback-Leibler information quantity of the two models  \citep{Burnham01}, i.e. how much more information is lost by describing the data as model $A$ rather than model $B$.  A lower $AIC$ value indicates a better model of the data and $\Delta AIC_{AB} \ge 2$ is considered significant. (Equation~\ref{deltaAIC} can be compared to the test statistic of a likelihood ratio $-2\ln (L_{a}/L_{b})=-2(\ln L_{a} - \ln L_{b})$ which approximately follows a $\chi^{2}$ distribution.)

Table~\ref{MainTable} shows the percentage of the $1000$ simulations for which there is a detection of significant effects in the spectrum due to photon-ALP mixing, i.e. including ALPs in the model provides a significant improvement to fitting the measured spectrum compared to not including them.  From here on we shall refer to a significant detection of photon-ALP mixing effects as the detection of a `signature'.

The model of the Milky Way's magnetic field is a larger influence on the results than the EBL model, although both are important.  With the Milky Way magnetic field model of Jansson \& Farror \cite{Jansson12} prospects for detecting a signature are very good with $>60\%$ chance for all $3$ objects.  Including the effects of photon-ALP mixing within the jet typically adds $5$-$10\%$ chance to the prospects of detecting a signature.  If the jet is assumed to be short, $\Delta Z=0.1\rm~kpc$, the flux boost due to Cotton-Mouton suppression typically adds $30\%$ or more to the chances of detecting a signature.

As can be seen, PKS~2155-304 is the most promising target, presumably due to its brightness.  Even with the more pessimistic Milky Way model in which only a fraction of $0.05$ of ALPs reconvert, detection prospects are always above $50\%$.

For 1ES~1101-232 the detection prospects are still good, again over $50\%$ in all the scenarios considered here.  PKS~2005-489 shows a wider variation in the prospects for detecting a signature.  Prospects are as low as $11\%$ in the most pessimistic scenario, but there is a $100\%$ chance of detecting a signature if $\Delta Z = 0.1\rm~kpc$.  This can be understood when we consider the object's relative closeness, in which case the additional $\gamma$-rays produced at the highest energies do not suffer much absorption from the EBL.

\begin{table*}
 \centering
  \footnotesize\setlength{\tabcolsep}{1.0pt}
 \begin{minipage}{140mm}
  \caption{Percentage of simulations where including photon-ALP conversion in the model fitted to the measured spectrum is a significant improvement.\label{MainTable}}
  \begin{tabular}{@{}lc|ccc|ccc@{}}
  \hline

 &  &  & Kneiske \& Dole & & & Francheshini et al. &  \\
Object & Reconversion & Conv & No Conv & $\Delta Z=$ & Conv& No Conv& $\Delta Z=$ \\
Name & Fraction & in Jet &in Jet & $0.1\rm~kpc$&in Jet&in Jet& $0.1\rm~kpc$ \\

\hline
1ES 1101-232 & 0.25 & 81\% & 73\% & 86\% & 93\% & 88\% & 79\% \\
              & 0.05 & 54\% & 54\% & 96\% & 63\% & 61\%  & 84\%\\
	      &&&&&\\
PKS 2155-304  & 0.60 & 100\% & 100\% & 100\% & 100\% & 100\% & 100\%\\
              & 0.05 & 76\% & 65\% & 100\% & 68\% & 54\% & 100\% \\
      	      &&&&&\\
PKS 2005-489  & 0.70 & 71\% & 60\% & 100\% & 91\% & 84\% & 100\% \\
              & 0.05 & 15\% & 11\% & 100\% & 25\% & 19\% & 100\% \\

\hline

 \end{tabular}
 \end{minipage}
 \end{table*}

\section{Summary}
If the hypothetical axions or axion-like particles (ALPs) exist, then photon-ALP oscillation would be possible in the presence of a magnetic field.  In this paper we consider the conversion of $\gamma$-ray photons to ALPs in the magnetic field of AGN jets and in the magnetic field external to AGN jets.  We then calculate the probability of the next generation $\gamma$-ray telescope CTA detecting the signature of photon-ALP mixing in the spectra of $3$ objects: 1ES~1101-232, PKS~2155-304 and PKS~2005-489.  Such signatures can be due to changes in the emitted spectra caused by photon-ALP mixing within the jet as well as ALPs mitigating the absorption effects of low energy extragalactic background light (EBL) photons on $\gamma$-rays.  The prospects for detecting the signature of photon-ALP mixing in the spectra of an object vary between $11\%$ and $100\%$ depending upon assumptions in the EBL model, the Milky Way magnetic field model, and the length of the jet.  The assumptions about the Milky Way magnetic field model appear to have a bigger impact on the results than the assumptions about the EBL model, although both are important.  The most consistently promising target is PKS~2155-304, which suggests that observations should target relatively bright, nearby objects as well as distant hard-spectrum sources.

Throughout this paper we have assumed an ALP mass of $10^{-8}\rm~eV$ and coupling constant of $5 \times 10^{-11} \rm~GeV^{-1}$.  In our jet model, photons are initially emitted near the base of the jet and propagate up to the jet termination.  The particle density and magnetic field strength decrease with distance from the emission region.

We find that, in general, due to the wide range of particle densities and magnetic field strengths the photon beam passes through as it travels up the jet, the effects of photon-ALP mixing preserve the spectral index of the $\gamma$-ray spectrum. Including the effects of photon-ALP mixing within the jet typically provide an additional $5$-$10\%$ chance of detecting a signature, compared with assuming that photon-ALP mixing occurs only outside the jet, e.g. in the magnetic field of the host galaxy cluster.  It also means that $\gamma$-rays emitted from blazars not hosted within a galaxy cluster will undergo photon-ALP mixing and may produce detectable signatures.

If the emission region is close to the termination of the jet, $\Delta Z = 0.1\rm~kpc$, then the prospects for detecting a signature see a large increase, especially in PKS~2005-489.  This is due to a sudden jump in $\gamma$-ray flux where Cotton-Mouton suppression occurs at around $1.7\rm~TeV$ in the object's rest frame.   This suggests a search strategy of looking at relatively nearby blazars where these additional $\gamma$-rays do not suffer much absorption with the EBL.  However, in general $\Delta Z$ would be expected to be much larger than $0.1\rm~kpc$.  Furthermore, Cotton-Mouton suppression could reasonably be assumed to occur at higher energies where photon numbers are poorer even if $\Delta Z \leq 0.1\rm~kpc$.  If the $\gamma $-ray emission originated in a small region close to the termination of the jet, for example due to magnetic reconnection, the magnetic field strength immediately outside the emission region might be small.  Similarly, even if the jet is disrupted close to the central black hole causing a small $\Delta Z$, small magnetic field strengths may be encountered in the disrupted material beyond the termination of the jet.  In these cases the lower magnetic field strengths would push the Cotton-Mouton suppression to higher energies.

Other effects not considered in this paper may degrade the prospects of detecting a signature, although not dramatically.  Firstly, the CTA's energy resolution will not be perfect, causing a smearing of the flux boost at high energies, which was not included in our simulations.  Secondly, the spectra of the objects in this paper were treated as simple power laws when in reality they may have curvature.  This would lower the flux at higher energies but again this would not be expected to have a large effect on the results, especially taking into account the range of spectral parameters considered in this paper.

As well as the $3$ targets in this paper, several other promising targets exist such as PKS~0301-243, 3C~279 and PKS~1424+240.  There are no compelling a priori reasons to assume that the EBL should be at its lower limits across a wide range of wavelengths or to assume that no photon-ALP mixing would occur within the jet.  Therefore, if ALPs exist with the parameters assumed here, we conclude that signatures are expected to be seen in the spectrum of several objects. Conversely, if the effects are not seen, limits can be placed on the ALP parameter space.

\acknowledgments

We would like to thank Emma Gardner, Jim Hinton, Richard White, Jonathan Davis, Thomas Armstrong and our colleagues in the CTA consortium for their most helpful contributions to this paper.  We would also like to thank Sam Nolan for his careful reading of the manuscript.  JH acknowledges funding from the UK STFC under grant ST/I505656/1.

\bibliographystyle{JHEP}
\bibliography{bib}{}

\providecommand{\href}[2]{#2}\begingroup\raggedright\begin{thebibliography}{10}

\bibitem{Gerasimova62}
N.~M. Gerasimova, A.~I. Nikishov, and I.~L. Rosenthal, {\it Interation of
  nuclei and photons of high energies with a thermal radiations in the
  universe},  {\em Journal of the Physical Society of Japan.} {\bf 17} (1962)
  Suppl. A--III.

\bibitem{Aharonian06}
F.~Aharonian et~al., {\it A low level of extragalactic background light as
  revealed by $\gamma$-rays from blazars},  {\em Nature} {\bf 440} (2006)
  1018--2021.

\bibitem{Lefa11}
E.~Lefa, F.~M. Riegger, and F.~Aharonian, {\it Formation of very hard gamma-ray
  spectra of blazars in leptonic models},  {\em ApJ} {\bf 740} (2011) 64--73.

\bibitem{Csaki03}
C.~Cs\'{a}ki, N.~Kaloper, M.~Peloso, and J.~Terning, {\it Super {GZK} photons
  from photon axion mixing},  {\em JCAP} {\bf 05} (2003) 005.

\bibitem{DeAngelis07}
A.~D. Angelis, M.~Roncadelli, and O.~Mansutti, {\it Evidence for a new light
  spin-zero boson from cosmological gamma-ray propagation?},  {\em Phys Rev D}
  {\bf 76} (2007) 121301(R).

\bibitem{Horns12}
D.~Horns et~al., {\it Hardening of {TeV} gamma spectrum of active galactic
  nuclei in galaxy clusters by conversions of photons into axionlike
  particles},  {\em Phys Rev D} {\bf 86} (2012) 075024.

\bibitem{Finke10}
J.~D. Finke, S.~Razzaque, and C.~D. Dermer, {\it Modeling the extragalactic
  background light from stars and dust},  {\em ApJs} {\bf 712} (2010) 238--249.

\bibitem{Weinberg78}
S.~Weinberg, {\it New light boson},  {\em Phys Rev Lett} {\bf 40} (1978)
  223--226.

\bibitem{Wilczek78}
F.~Wilczek, {\it Problem of strong {P} and {T} invariance in the presence of
  instantons},  {\em Phys Rev Lett} {\bf 40} (1978) 279--282.

\bibitem{Peccei77}
R.~D. Peccei and H.~R. Quinn, {\it {CP} conservation in the presence of
  pseudoparticles},  {\em Phys Rev Lett} {\bf 38} (1977) 1440--1443.

\bibitem{Isern10}
J.~Isern, E.~Garc\'{i}a, L.~G. Althaus, and A.~H. C\'{o}rsico, {\it Axions and
  the pulsation periods of variable white dwarfs revisited},  {\em A\&A} {\bf
  512} (2010) A86.

\bibitem{Park12}
C.-G. Park, J.~c.~Hwang, and H.~Noh, {\it Axion as a cold dark matter
  candidate: Low-mass case},  {\em Phys Rev Lett} {\bf 86} (2012) 083535.

\bibitem{Simet08}
M.~Simet, D.~Hooper, and P.~Serpico, {\it Milky way as a kiloparsec-scale
  axionscope},  {\em Phys Rev D} {\bf 77} (2008) 063001.

\bibitem{HornsMeyer12}
D.~Horns and M.~Meyer, {\it Indications for a pair-production anomaly from the
  propagation of {VHE} gamma-rays},  {\em JCAP} {\bf 02(2012)} (2012) 33.

\bibitem{Sanchez-Conde09}
M.~A. S\'{a}nchez-Conde et~al., {\it Hints of the existence of
  {Axion-Like-Particles} from the gamma-ray spectra of cosmological sources},
  {\em Phys Rev D} {\bf 79} (2009) 123511.

\bibitem{Hochmuth07}
K.~A. Hochmuth and G.~Sigl, {\it Effects of axion-photon mixing on gamma-ray
  spectra from magnetized astrophysical sources},  {\em Phys Rev D} {\bf 76}
  (2007) 123011.

\bibitem{Tavecchio12}
F.~Tavecchio et~al., {\it Evidence for an axion-like particle from {PKS} 1222 +
  216?},  {\em Phys Rev D} {\bf 86} (2012) 085036.

\bibitem{Mena13}
O.~Mena and S.~Razzaque, {\it Hints of an axion-like particle mixing in the
  {GeV} gamma-ray blazar data?},  {\em JCAP} {\bf 11(2013)} (2013) 023.

\bibitem{Stawarz03}
{\L}.~Stawarz, M.~Sikora, and M.~Ostrowski, {\it High-energy gamma rays from
  {FR~I} jets},  {\em ApJ} {\bf 597} (2003) 186--201.

\bibitem{Blundell01}
K.~M. Blundell and S.~Rawlings, {\it The optically powerful quasar {E}1821+643
  is associated with a 300 kiloparsec-scale {FR}~i radio structure},  {\em ApJ}
  {\bf 562} (2001) L5--L8.

\bibitem{Cheung07}
C.~C. Cheung, D.~E. Harris, and {\L}.~Stawarz, {\it Superluminal radio features
  in the {M87} jet and the site of flaring {TeV} gamma-ray emission},  {\em
  ApJ} {\bf 663} (2007) L65--L68.

\bibitem{Tavecchio13}
F.~Tavecchio et~al., {\it The far emission region of the gamma-ray blazar {PKS
  B1424-418}},  {\em MNRAS} {\bf 435} (2013) L24--L28.

\bibitem{Aharonian07}
F.~Aharonian et~al., {\it Detection of {VHE} gamma-ray emission from the
  distant blazar {1ES} 1101-232 with {HESS} and broadband characterisation},
  {\em A\&A} {\bf 470} (2007) 475--489.

\bibitem{Ghisellini10}
G.~Ghisellini et~al., {\it General physical properties of bright {F}ermi
  blazars},  {\em MNRAS} {\bf 402} (2010) 497--518.

\bibitem{Blandford79}
R.~D. Blandford and A.~K\"onigl, {\it Relatavistic jets as compact radio
  sources},  {\em ApJ} {\bf 232} (1979) 34--48.

\bibitem{Heinz03}
S.~Heinz and R.~A. Sunyaev, {\it The non-linear dependence of flux on black
  hole mass and accretion rate in core dominated jets},  {\em MNRAS} {\bf 343}
  (2003) L59.

\bibitem{Falcke95}
H.~Falcke and P.~L. Biermann, {\it The jet-disk symbiosis {I.} radio to x-ray
  emission models for quasars},  {\em A\&A} {\bf 293} (1995) 665.

\bibitem{Gardner13}
E.~Gardner and C.~Done, {\it Jets and the accretion flow in low luminosity
  black holes},  {\em MNRAS} {\bf 434} (2013) 3454.

\bibitem{Katarzynski01}
K.~Katarzynski, H.~Sol, and A.~Kus, {\it The multifrequency emission of {Mrk}
  501. from radio to {TeV} gamma-rays},  {\em A\&A} {\bf 367} (2001) 809--825.

\bibitem{Lyubarsky10}
Y.~Lyubarsky, {\it A new mechanism for dissipation of alternating fields in
  {P}oynting-dominated outflows},  {\em ApJ} {\bf 725} (2010) L234--L238.

\bibitem{Raffelt88}
G.~Raffelt and L.~Stodolsky, {\it Mixing of the photon with low-mass
  particles},  {\em Phys Rev D} {\bf 37} (1988) 1237--1249.

\bibitem{Andriamonje07}
S.~Andriamonje, {\it An improved limit on the axion photon coupling from the
  {CAST} experiment},  {\em JCAP} {\bf 04(2007)} (2007) 10.

\bibitem{arXivnote}
P.~Brun and D.~Wouters, {\it Reply to comment on "{I}rregularity in gamma ray
  source spectra as a signature of axionlike particles"},  {\em arxiV} {\bf
  1305} (2013) 4098v1.

\bibitem{Ganguly09}
A.~K. Ganguly, P.~Jain, and S.~Mandal, {\it Photon and axion oscillation in a
  magnetized medium: A general treatment},  {\em Phys Rev D} {\bf 79} (2009)
  11501.

\bibitem{DeAngelis08}
A.~D. Angelis, O.~Mansutti, and M.~Roncadelli, {\it Axion-like particles,
  cosmic magnetic ﬁelds and gamma-ray astrophysics},  {\em Phys Lett B} {\bf
  659} (2008) 847--855.

\bibitem{Acharya13}
B.~S. Acharya et~al., {\it Introducing the {CTA} concept},  {\em Astropart
  Physics} {\bf 43} (2013) 3--18.

\bibitem{Bernlohr13}
K.~Bernl\"ohr et~al., {\it {Monte Carlo} design studies for the {Cherenkov
  Telescope Array}},  {\em Astropart Physics} {\bf 43} (2013) 171--188.

\bibitem{DiPierro11}
F.~D. Pierro et~al., {\it Performance studies of the {CTA} observatory},  {\em
  Proc. of 32nd Int. Cosmic Ray Conference.} (2011).

\bibitem{Acero10}
F.~Acero et~al., {\it {PKS} 2005−489 at {VHE}: four years of monitoring with
  {HESS} and simultaneous multi-wavelength observations},  {\em A\&A} {\bf 511}
  (2010) A52.

\bibitem{Abramowski10}
A.~Abramowski et~al., {\it {VHE} gamma-ray emission of {PKS} 2155–304:
  spectral and temporal variability},  {\em A\&A} {\bf 520} (2010) A83.

\bibitem{Jansson12}
R.~Jansson and G.~R. Farrar, {\it A new model of the galactic magnetic field},
  {\em ApJ} {\bf 757} (2012) 14--26.

\bibitem{Pshirkov11}
M.~S. Pshirkov et~al., {\it Deriving the global structure of the galactic
  magnetic field from {F}araday rotation measures of extragalactic sources},
  {\em ApJ} {\bf 738} (2011) 192--205.

\bibitem{Kneiske10}
T.~M. Kneiske and H.~Dole, {\it A lower-limit flux for the extragalactic
  background light},  {\em A\&A} {\bf 515} (2010) A19.

\bibitem{Franceschini08}
A.~Franceschini, G.~Rodighiero, and M.~Vaccari, {\it Extragalactic
  optical-infrared background radiation, its time evolution and the cosmic
  photon-photon opacity},  {\em A\&A} {\bf 487} (2008) 837--852.

\bibitem{Burnham01}
K.~P. Burnham and D.~R. Anderson, {\it {Kullback-Leibler} information as a
  basis for strong inference in ecological studies},  {\em Wildlife Research}
  {\bf 29} (2001) 111--119.

\end{thebibliography}\endgroup

\clearpage

\clearpage
\clearpage

\clearpage

\clearpage

\clearpage

\end{document}